\documentclass[lettersize,journal]{IEEEtran}
\IEEEoverridecommandlockouts

\usepackage{amsmath,amssymb,amsfonts,amsthm}
\usepackage[scr=euler]{mathalpha}
\usepackage{enumitem}
\usepackage{algorithm}
\usepackage{algpseudocode}
\usepackage{mathtools}
\usepackage{cleveref}
\crefname{figure}{Fig.}{Fig.}
\crefformat{equation}{(#2#1#3)}
\usepackage{caption}
\usepackage{dsfont}
\usepackage{subfig}    
\usepackage{stfloats}

\usepackage{graphicx}
\usepackage{textcomp}
\usepackage{xcolor}

\usepackage{cite}
\usepackage[acronym]{glossaries}
\usepackage{tabularx}

\def\BibTeX{{\rm B\kern-.05em{\sc i\kern-.025em b}\kern-.08em
    T\kern-.1667em\lower.7ex\hbox{E}\kern-.125emX}}

\theoremstyle{definition}
\newtheorem{definition}{Definition}

\DeclareMathOperator*{\argmax}{arg\,max}

\newacronym{In-F}{In-F}{In-Factory}
\newacronym{6G}{6G}{sixth generation}
\newacronym{5G}{5G}{fifth generation}
\newacronym{SN}{SN}{sub-network}
\newacronym{HRLLC}{HRLLC}{hyper-reliable, low-latency communication}
\newacronym{URLLC}{URLLC}{ultra-reliable, low-latency communication}
\newacronym{MCS}{MCS}{modulation and coding scheme}
\newacronym{LA}{LA}{Link adaptation}
\newacronym{CQI}{CQI}{channel quality indicator}
\newacronym{LSTM}{LSTM}{Long short-term memory}
\newacronym{SINR}{SINR}{signal to interference plus noise ratio}
\newacronym{ACK}{ACK}{acknowledgement}
\newacronym{NACK}{NACK}{non-acknowledgement}
\newacronym{BLER}{BLER}{block error rate}
\newacronym{OLLA}{OLLA}{outer-loop link adaptation}
\newacronym{LOS}{LOS}{line-of-sight}
\newacronym{NLOS}{NLOS}{non-line-of-sight}
\newacronym{IIR}{IIR}{infinite impulse response}
\newacronym{eMBB}{eMBB}{enhanced mobile broadband}
\newacronym{TTI}{TTI}{transmission time interval}
\newacronym{IPV}{IPV}{interference power value}
\newacronym{pdf}{pdf}{probability density function}
\newacronym{MO}{MO}{modulation order}
\newacronym{CR}{CR}{coding rate}
\newacronym{ECDF}{ECDF}{Empirical cumulative distribution function}
\newacronym{EKF}{EKF}{extended Kalman filter}
\newacronym{SA}{SA}{sensor-actuator pair}
\newacronym{EESM}{EESM}{exponential effective signal-to noise-ratio mapping}
\newacronym{ESM}{ESM}{effective SNR mapping}
\newacronym{MIESM}{MIESM}{mutual information effective SNR mapping}
\newacronym{CSI-RS}{CSI-RS}{channel state information-reference Signal}
\newacronym{MSE}{MSE}{mean square error}
\newacronym{IM}{IM}{Interference management}
\newacronym{AP}{AP}{access point}
\newacronym{UE}{UE}{user equipment}
\newacronym{CSI}{CSI}{channel state information}
\newacronym{MAC}{MAC}{medium access control}
\newacronym{INR}{INR}{Interference-to-noise-ratio}
\newacronym{3GPP}{3GPP}{third generation partnership project}
\newacronym{RDMM}{RDMM}{random directional mobility model}
\newacronym{InF-DL}{InF-DL}{indoor factory with dense clutter and low base station}
\newacronym{SNR}{SNR}{signal to noise ratio}
\newacronym{DSSM}{DSSM}{dynamic state space model}
\newacronym{vDSSM}{vDSSM}{vector dynamic state space model}
\newacronym{DL}{DL}{downlink}
\newacronym{TDD}{TDD}{time division duplexing}
\newacronym{eSNR}{eSNR}{effective signal-to-noise ratio}
\newacronym{RRM}{RRM}{radio resource management}
\newacronym{RAE}{RAE}{relative absolute error}
\newacronym{GRU}{GRU}{gated recurrent unit}
\newacronym{GPR}{GPR}{Gaussian Process Regression}
\newacronym{TPR}{TPR}{Student-t Process Regression}
\newacronym{KL}{KL}{Kullback-Leibler}
\newacronym{SPTPR}{SPTPR}{sparse Student-t process regression}
\newacronym{MUKF}{MUKF}{modified unscented Kalman filter}
\newacronym{HARQ}{HARQ}{hybrid automatic repeat request}
\newacronym{UT}{UT}{unscented transform}
\newacronym{LTE}{LTE}{long term evolution}
\newacronym{MA}{MA}{moving-average}
\newacronym{UKF}{UKF}{unscented Kalman filter}
\newacronym{SE}{SE}{spectral efficiency}
\newacronym{Tx}{Tx}{transmission}
\newacronym{SNC}{SNC}{SN controller}

\title{CQI-Based Interference Prediction for Link Adaptation in Industrial Sub-networks}
\author{\IEEEauthorblockN{Pramesh Gautam, Ravi Sharan B A G, Paolo Baracca,  Carsten Bockelmann, Thorsten Wild, Armin Dekorsy}
    \thanks{
		This work is supported by the German Federal Ministry of Research, Technology and Space~(BMFTR) under the grants of 16KIS077K, 16KISK109~(6G-ANNA) and 16KISK016~(Open6GHub). Pramesh Gautam, Carsten Bockelmann, and  Armin Dekorsy are with the Department of Communications Engineering, University of Bremen,  Germany (Email: \{gautam, bockelmann, dekorsy\}@ant.uni-bremen.de).
        Paolo Baracca is with Nokia Standards,  Germany. Email: \{paolo.baracca\}@nokia.com. Ravi Sharan B A G, and Thorsten Wild are with Nokia Bell Labs Stuttgart, Germany. Email: \{ravi.sharan, thorsten.wild\}@nokia-bell-labs.com.
}%

}
\usepackage[firstpage=true]{background}
\SetBgContents{\textcolor{black}{\footnotesize This work has been submitted to the IEEE for possible publication. Copyright may be transferred without notice, after which this version may no longer be accessible.}}
\SetBgPosition{current page.south}
\SetBgVshift{0.50cm}
\SetBgOpacity{1.0}
\SetBgAngle{0.0}
\SetBgScale{1.0}
\begin{document}
\maketitle
\begin{abstract}
We propose a novel interference prediction scheme to improve link adaptation~(LA) in densely deployed industrial sub-networks~(SNs) with high-reliability and low-latency communication~(HRLLC) requirements. The proposed method aims to improve the LA framework by predicting and leveraging the heavy-tailed interference probability density function~(pdf). Interference is modeled as a latent vector of available channel quality indicator~(CQI), using a vector discrete-time state-space model (vDSSM) at the SN controller, where the CQI is subjected to compression, quantization, and delay-induced errors. To robustly estimate interference power values under these impairments, we employ a low-complexity, outlier-robust, sparse Student-t process regression~(SPTPR) method. This is integrated into a modified unscented Kalman filter, which recursively refines predicted interference using CQI, enabling accurate estimation and compensating protocol feedback delays—crucial for accurate LA. Numerical results show that the proposed method achieves over 10× lower complexity compared to a similar non-parametric baseline. It also maintains a BLER below the 90th percentile target of $10^{-6}$ while delivering performance comparable to a state-of-the-art supervised technique using only CQI reports.

\end{abstract}

\section{Introduction}
\label{sec:introduction}
\gls{LA} is a fundamental feature in wireless communication systems, enabling the selection of \gls{MCS} to optimize data rates and ensure reliability under varying channel conditions. The \gls{LA} problem has been effectively managed in services like enhanced mobile broadband~(eMBB) with relaxed latency and reliability requirements~\cite{brighente2020interference}. However, this method becomes inadequate in emerging scenarios such as industrial \glspl{SN}, which require closed-loop control supported by short-range communications under \gls{HRLLC} traffic with strict latency (0.1–1 ms) and reliability requirements~\cite{berardinelli2021extreme}. To achieve extremely low \gls{BLER}, smaller \gls{OLLA} step sizes are needed. This results in slow convergence \cite{10024770}, further exacerbated by significant \gls{SINR} fluctuations caused by co-subband interference stemming from limited subbands, dense deployments, and mobility. Moreover, low latency requirements of \glspl{SN} severely limit packet retransmissions, making \gls{ACK}/non-\gls{ACK} feedback less practical and further complicating \gls{LA}. To address these limitations, we propose a predictive \gls{LA} scheme tailored for interference-prone \glspl{SN}, enabling single transmissions under stringent latency and reliability constraints by efficiently leveraging the underlying \gls{pdf} of the inter-SN \gls{IPV}. \gls{IPV} prediction has been widely studied, but as this work focuses on \gls{SNC}-side (access point~(AP)-side) \gls{LA} enhancement, we limit discussion to studies leveraging predicted \gls{IPV} for \gls{LA} \cite{pramesh_kalmam}. In \cite{pocovi2017mac}, the authors proposed a \gls{UE}-side method (as \gls{SA} in this work) by incorporating historical interference estimates into \gls{SINR} using a first-order low-pass filter. Brighente et al. \cite{brighente2020interference} proposed low-complexity UE-side prediction with Gaussian kernel methods, though the Gaussian assumption limits generalization due to interference's heavy-tailed distribution. Wei et al. \cite{wei2024joint} combine seasonal and trend decomposition using locally estimated scatterplot smoothing and require a \gls{GRU}-based prediction. They require three GRU models per \gls{SA}, raising scalability concerns for low-form-factor SNs with multiple \glspl{SA}. Although not directly an \gls{IPV} prediction scheme, the authors in \cite{xu2013improving} propose an \gls{AP} side Wiener filtering based \gls{CQI} prediction for \gls{LTE}.
\begin{figure}
    \centering
    \includegraphics[width=0.8\linewidth]{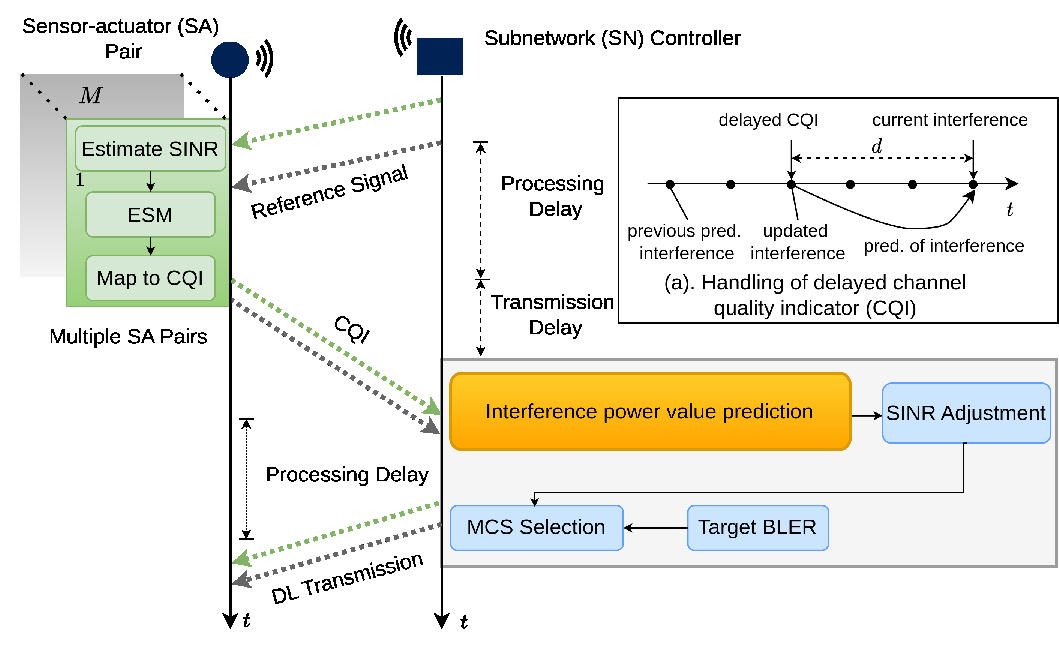}
        \vspace{-0.10in}
    \caption{Interference Prediction for \gls{LA} from time-delayed CQI.}
    \label{fig:proposed_LA}
    \vspace{-0.25in}
\end{figure}
In our recent work, we proposed a scalar \gls{DSSM} with linear approximation for \gls{LA} in \glspl{SN} for a single \gls{SA} from \gls{CQI} solved with Gaussian assumption \cite{pramesh_kalmam}. We extend this framework by proposing a novel predictive \gls{LA} framework with the ability \textit{to compensate for delays associated with outdated \gls{CQI}} using a fully data-driven approach for \gls{SN} with \textit{multiple \glspl{SA}}. It captures heavy-tailed interference by leveraging \gls{SPTPR} for functional inference and covariance estimation, and recursively incorporates delayed CQI through a Kalman filter to enable more timely and accurate \gls{LA} (see Fig.~\ref{fig:proposed_LA}). To the best of our knowledge, this is the first \gls{vDSSM}-based approach for \gls{LA}, enriched with \gls{IPV} prediction beyond Gaussian assumptions in \glspl{SN}, and incorporating \gls{3GPP} signaling errors as well as inherent protocol delays under \gls{HRLLC} constraints.

\section{System Model}
\label{sec:systemmodel}
Let \(\mathscr{N} := \{1, 2, \ldots, N\}\), with \(|\mathscr{N}| = N\), denote the set of \glspl{SN} moving at a constant velocity \(v\) in random directions that are uniformly distributed within a confined area \(\mathscr{A}\). Each \gls{SN} \(n \in \mathscr{N}\) comprises a single \gls{SNC} serving a set \(\mathscr{M}_{n} := \{m_1, m_2, \ldots, m_M\},~|\mathscr{M}_{n}| = M\), of \glspl{SA} that enable closed-loop control communication. The total bandwidth \(B\) is partitioned into \(\Omega\) sub-bands (\(\Omega \ll N\)), with each \gls{SN} operating in \gls{TDD} mode and assigned a dedicated sub-band to support intra-\gls{SN} communication. Each \gls{Tx} cycle, denoted by \(t \in \mathbb{N}\), has a frame duration \(T_F\), which is divided into uplink and downlink phases of durations \(T_{\text{UL}}\) and \(T_{\text{DL}}\), respectively. In this work, we focus on the downlink, where \(T_{\text{DL}}\) is further split into \(N_{\text{sl}}\) slots of duration \(T_s\), such that \(T_{\text{DL}} = N_{\text{sl}} \cdot T_s\), where \(T_s\) is assumed to be a multiple of an OFDM symbol to comply with \gls{3GPP} specifications. A slot of duration \(T_s\) is allocated to a single \gls{SA} for downlink transmission, denoted by \(m_{\tau}\), representing the \(m\)-th \gls{SA} active during the \(\tau\)-th slot of the \(t\)-th \gls{Tx} cycle.  Although orthogonal slot allocation minimizes intra-\gls{SN} interference, dense deployments and the need for large subband bandwidths for inter-\gls{SN} communication result in substantial inter-\gls{SN} interference \cite{pramesh_kalmam}. To perform \gls{LA} in downlink for each \gls{SN}, we adopt the standard feedback process, where the \gls{SNC} transmits reference signals allowing each \gls{SA} to estimate the \gls{SINR} as \(\tilde{\xi}_{m_{\tau} n}(t) = S_{m_{\tau} n}(t) / (\iota_{m_{\tau} n}(t)+ \sigma_w^2)\), with \(S_{m_{\tau} n}(t)\), \(\iota_{m_{\tau} n}(t)\), and \(\sigma_w^2\) denoting signal power, inter-\gls{SN} \gls{IPV}, and noise power, respectively. The estimated SINR is converted into an \gls{eSNR} using an \gls{ESM} or \gls{MIESM} mechanism, mapped to a \gls{CQI} via predefined look-up tables, and reported to the \gls{SNC} either periodically or periodically \cite{xu2013improving,10024770}. At each \gls{Tx} cycle \(t\), the \gls{SNC} collects wideband \gls{CQI} from all scheduled \glspl{SA} and stores it as \(\mathbf{y}_n(t)\). Since the \gls{SNC} and \glspl{SA} are deployed in fixed positions at short range, the \gls{CQI} variations primarily reflect  \gls{IPV} dynamics. Slot-level \gls{IPV} for \(m_{\tau} n\) arises from the set \(\mathcal{C}_{m_{\tau} n} = \bigcup_{n' \in \hat{\mathscr{N}} \setminus \{n\}} m_{\tau} n'\), where \(\hat{\mathscr{N}} \subset \mathscr{N}\) includes \glspl{SN} scheduled on the same sub-band during slot \(\tau\). When modeling slot-level co-subband \gls{IPV} using the spatially consistent \gls{3GPP} channel model—which incorporates both \gls{LOS} and \gls{NLOS} components for dynamically moving \glspl{SN} \cite{pramesh_kalmam}—we observe a heavy-tailed \gls{IPV} distribution. This arises from deep fading, finite blocklength transmissions, sporadic traffic, and a limited number of interferers combined with industrial shadowing, consistent with the findings in \cite{clavier2020experimental}.


\section{Link Adaptation Problem}
\label{sec.probformulation}
In this work, we consider the \gls{LA} problem, which includes selecting an appropriate \gls{MCS} under reliability constraints. Specifically, the \gls{SNC} selects the \gls{MCS} based on wideband \gls{CQI} reports from its associated \glspl{SA}. The proposed \gls{LA} problem for the \(n\)-th \gls{SN} with multi-\glspl{SA} at \gls{Tx} cycle \(t\) is:
\begin{align}
\label{eq:laprob}
\boldsymbol{\lambda}_{n}^*(t) =   \argmax_{\boldsymbol{\lambda}_{n}(t)} \mathbf{r}_{n}(\boldsymbol{\lambda}_{n}(t) | \hat{\boldsymbol{\xi}}_{n}(t)) \nonumber \\
\text{s.t.} \quad \boldsymbol{\varepsilon}(\boldsymbol{\lambda}_{n}(t) | \hat{\boldsymbol{\xi}}_{n}(t)) \preceq \Bar{\boldsymbol{\varepsilon}}.
\end{align}
where, $\boldsymbol{\lambda}_{n}^*(t) := [{\lambda}_{m_{1} n}^*(t) , \dotsc, {\lambda}_{m_{M} n}^*(t) ] \in \zeta$ is the optimal \gls{MCS} vector with $\zeta$ being the set of all feasible \gls{MCS}. The quantity $\mathbf{r}_n(\cdot)= [{r}_{m_{1} n}(t) , \dotsc, {r}_{m_{M} n}(t) ]$ denotes the achievable spectral efficiency based on selected \gls{MCS} values for $n^{\text{th}}$ \gls{SN}. Finally $\boldsymbol{\varepsilon}(t)$ and $\Bar{\boldsymbol{\varepsilon}}$ represent the achieved \gls{BLER} and target \gls{BLER}, respectively. The main challenges in solving~\eqref{eq:laprob} at the \gls{SNC} stems from inaccurate interference representation due to compression losses in \gls{ESM}, limited \gls{CQI} quantization order, and outdated \gls{CQI} at the \gls{SNC} caused by protocol delays. It can be incorporated into~\eqref{eq:laprob} by: (i) estimating it at the \gls{SA} and sending it to the \gls{SNC} with additional signaling overhead; (ii) predicting it locally in the \gls{SA} and sending it to \gls{SNC} with additional signaling, computation, and energy demands; or (iii) predicting it at the \gls{SNC} using available \gls{CQI}. We adopt the third approach following 3GPP~(based on \cite{3gpp2018physical}), without incurring additional signaling or energy costs at the \gls{SA}.

\section{Dynamic State Space Modeling}
\label{sec:dssmcharacterization}

We propose a \gls{vDSSM} at the \gls{SNC} to estimate heavy-tailed \gls{IPV} vector. These are treated as latent variables, and the core task is to infer them from available channel state information—i.e., \gls{CQI} reports from associated \glspl{SA}. Specifically, the \gls{vDSSM} at the \(n^{\text{th}}\) \gls{SNC} serving \(M\) \glspl{SA} is given by:\begin{subequations}
\label{eq:dssm1}
\begin{align}
    &\boldsymbol{\iota}_n(t) := F(\boldsymbol{\iota}_n(t-1), \boldsymbol{\beta}(t)), \label{eq:dssm1a} \\
    &\mathbf{y}_{n}(t) := G(\boldsymbol{\iota}_n(t), \mathbf{u}(t)).\label{eq:dssm1b}
\end{align}
\end{subequations}
The process model in~\eqref{eq:dssm1a} describes the evolution of the \gls{IPV} over \gls{Tx} cycles, treated as the latent variable in the \gls{vDSSM}. The measurement model in~\eqref{eq:dssm1b} relates the outdated \gls{CQI} report of the \(n\)-th \gls{SN}, \(\mathbf{y}_n(t):= [y_{m_1 n}(t), \dotsc, y_{m_M n}(t)]^T \), to the \gls{IPV} \(\boldsymbol{\iota}_n(t):= [\iota_{m_1 n}(t), \dotsc, \iota_{m_M n}(t)]^T \in \mathbb{R}^{1 \times M}\) via unknown functions \(G(\cdot)\) and \(F(\cdot)\), which account for quantization, \gls{SINR} errors, and protocol delay. The symbols $\boldsymbol{\beta}(t)$ and  $\boldsymbol{\nu}(t)$ represent the measurement and process noise, respectively. The time-varying positions and temporal correlations render closed-form modeling of \gls{pdf} intractable, and analytical models derived for specific scenarios lack the flexibility to generalize for highly dynamic \gls{SN}\cite{gautam2023co}. We adopt a data-driven strategy to learn the unknown process and measurement models directly from observed \gls{CQI}. Specifically, we use non-parametric inference of \(G(\cdot)\) and \(F(\cdot)\) to model \(p(\mathbf{y}_n(t) | \boldsymbol{\iota}_n(t))\) with covariance \(\mathbf{C}_G\), and \(p(\boldsymbol{\iota}_n(t)|\boldsymbol{\iota}_n(t-1))\)  with covariance \(\mathbf{C}_F\) to capture dynamics across \gls{Tx} cycles. Importantly, \gls{vDSSM} in \cref{eq:dssm1} enables recursive refinement of the \gls{IPV} prediction with each new \gls{CQI} observation, in contrast to \cite{brighente2020interference}, which relies on a fixed training set, as detailed in the next section.
\section{Sparse Student-t Process regression}
\label{sec:ekfprediction}
We propose to utilize a Bayesian framework with a non-parametric probabilistic model, i.e., \gls{GPR}-based technique to learn the process and measurement models \(F(\cdot)\) and \(G(\cdot)\), along with their noise covariances \(\mathbf{C}_F\) and \(\mathbf{C}_G\) \cite{rasmussen2010gaussian}. \gls{GPR} faces two key challenges in \gls{IPV} prediction for industrial \gls{SN}: (i) \gls{IPV} often deviates from Gaussian assumption, whose rapidly decaying tails may under-represent rare interference events—becoming more severe under \gls{LA} optimization constraint in \cref{eq:laprob} \cite{brighente2020interference,clavier2020experimental}; and (ii) learning highly dynamic interference requires more data, making \gls{GPR} computationally prohibitive due to its \(\mathcal{O}(L^3)\) complexity from matrix inversion, where $L$ is training instances. To address these issues, we adopt \glsentryfull{SPTPR} with a heavy-tailed prior within a Bayesian framework. The Student-t distribution (by dropping subscript $\tau$ and $n$ for ease of notation) is defined as:
\vspace{-0.1in}
\begin{definition}
  The Student-t distribution with \( \nu \) degrees of freedom and \( \theta \) scale parameter for $m$-th SA is defined as ~\cite{shah2014student}
  \begin{equation}\label{eq:std_t}
p(\boldsymbol{\gamma}_{m}|\mathbf{f}_{m}) = \frac{\Gamma\left(\frac{\nu + 1}{2}\right)}{\Gamma\left(\frac{\nu}{2}\right) \sqrt{\nu\pi}\theta} \left( 1 + \frac{(\boldsymbol{\gamma}_{m} - \mathbf{f}_{m})^2}{\nu\theta^2} \right)^{-\frac{\nu+1}{2}},
\end{equation}
where $\Gamma(\cdot)$ is the gamma function, $\mathbf{f}_{m} = [f_{m}({\iota}_{m}(t))]$ and $\boldsymbol{\gamma}_{m} = [y_m(t)] \in \mathbb{R}^{1 \times L}$ are the function and corresponding observations over discrete \gls{Tx} cycles $t = 1, \dots, L$.
\end{definition}
Since the Student-t distribution is consistent under marginalization, \gls{TPR} can be defined using Kolmogorov’s consistency theorem as follows~\cite{xu2024sparse}:
\begin{definition}
  A function \( f_{m}(\cdot) \) is a Student-t process with parameter \( \nu > 2 \), mean function \( \mathbf{\mu}(\cdot) \), and kernel \( k(\cdot) \), the finite collection of function values has a joint multivariate Student-t distribution with scale \( \nu \), mean \( \mathbf{\mu}(\cdot) \), and covariance \( K \), where \( K_{t,t'} = k({\iota}_{m}(t), {\iota}_{m}(t')) \) is represented as \( \mathbf{f}_{m} \sim \mathcal{TP}(\nu, \mathbf{\mu}, K) \).
\end{definition}
TPR effectively models heavy-tailed interference and remains robust to outliers, making it suitable for sporadic traffic scenarios \cite{shah2014student}. In the context of interference prediction in ultra-dense deployments, \gls{IPV} may not exhibit inherent sparsity. To ensure computational scalability, we employ \gls{SPTPR} w.r.t. training data, which efficiently approximates the full posterior by selecting \(L_E \ll L\) inducing points from \(L\) training samples \(<\boldsymbol{\delta}_{m}, \boldsymbol{\gamma}_{m}>\) with $L$ sequential instances of $\iota_m$ and $y_m$ respectively, thereby capturing sufficient statistics for inference. Specifically, we choose \(L_E \ll L\) inducing points, denoted as \(\mathbf{Z}_{m} \in R^{1 \times L_{E}}\), with corresponding function values \(\mathbf{f}_{\mathbf{Z}_m} := [f_{\mathbf{Z}}(\mathbf{Z}_{m}(e)) \mid 1 \leq e \leq L_E]\). For simplicity, we denote \(\mathbf{f}_{m}\) instead of \(\mathbf{f}_{m}(\cdot)\). The conditional distribution \(p(\mathbf{f}_{m} \mid \mathbf{f}_{\mathbf{Z}_{m}})\) is defined as~\cite{shah2014student}:
\vspace{-0.1in}
\begin{align}
    p(\mathbf{f}_{m}| \mathbf{f}_{\mathbf{Z}_{m}}) =  \mathcal{SPT}\left( \nu + L_{E}, \mu, \frac{\nu + \eta - 2}{\nu + L_{E} - 2} \times \Xi \right),
\end{align}

where mean $\mu = K(\boldsymbol{\delta}_{m},{\mathbf{Z}_{m}}) [K({\mathbf{Z}_{m}},{\mathbf{Z}_{m}})+\sigma^{2}\boldsymbol{I}]^{-1} \mathbf{f}_{\mathbf{Z}_{m}}$,  $\eta = \mathbf{f}_{\mathbf{Z}_{m}}^T [K({\mathbf{Z}_{m}},{\mathbf{Z}_{m}})+\sigma^{2}\boldsymbol{I}]^{-1} \mathbf{f}_{\mathbf{Z}_{m}}$ and covariance $\Xi = K(\boldsymbol{\delta}_{m},{\mathbf{Z}_{m}}) - K(\boldsymbol{\delta}_{m},{\mathbf{Z}_{m}}) [K({\mathbf{Z}_{m}},{\mathbf{Z}_{m}})+\sigma^{2}\boldsymbol{I}]^{-1} K({\mathbf{Z}_{m}},\boldsymbol{\delta}_{m})$. This conditional distribution $p(\mathbf{f}_{m}| \mathbf{f}_{\mathbf{Z}_{m}})$ enables us to introduce the inducing points to compute the posterior distribution as follows:
\vspace{-0.1in}
\begin{equation}
    p(\mathbf{f}_{\mathbf{Z}_{m}}| \mathbf{f}_{m},\boldsymbol{\gamma}_{m}) =  \dfrac{p(\boldsymbol{\gamma}_{m}|\mathbf{f}_{m}) p(\mathbf{f}_{m}|\mathbf{f}_{\mathbf{Z}_{m}}) p(\mathbf{f}_{\mathbf{Z}_{m}})}{\int p(\boldsymbol{\gamma}_{m}|\mathbf{f}_{m}) p(\mathbf{f}_{m}|\mathbf{f}_{\mathbf{Z}_{m}}) p(\mathbf{f}_{\mathbf{Z}_{m}}) d\mathbf{f}_{\mathbf{Z}_{m}}}.
\end{equation}
To approximate the intractable posterior distribution \( p(\mathbf{f}_{\mathbf{Z}_{m}}|\mathbf{f}_{m},\boldsymbol{\gamma}_{m}) \), we introduce a variational distribution \( g(\mathbf{f}_{\mathbf{Z}_{m}}) \) with the variational parameter, such that the approximate posterior closely resembles the true posterior. To determine the variational parameters in \( g(\mathbf{f}_{\mathbf{Z}_{m}}) \), we minimize the negative \gls{KL} divergence between the true and variational posteriors using the reparameterization technique as follows:
\vspace{-0.2in}
\begin{multline} \label{eq:elbo}
    \log(p(\boldsymbol{\gamma}_{m})) =  \overbrace{\mathbb{E}_{p(\mathbf{f}_{m}|{\mathbf{f}_{\mathbf{Z}_{m}}}) \cdot g({\mathbf{f}_{\mathbf{Z}_{m}}})}[\log p(\boldsymbol{\gamma}_{m}|\mathbf{f}_{m})]}^{\text{Expected Likelihood function}} - \\\underbrace{KL(g({\mathbf{f}_{\mathbf{Z}_{m}}})||p({\mathbf{f}_{\mathbf{Z}_{m}}}))}_{\text{KL Regularization}}.
\end{multline}
Optimal hyperparameters are obtained via supervised training for the measurement model; the dynamic model is omitted for brevity but derived analogously. A trained \gls{SPTPR} approximates both measurement and dynamic models \cref{eq:dssm}, reducing complexity from $\mathcal{O}(L^{3})$ to $\mathcal{O}(L_{E}^{2} \cdot L)$.
\section{Proposed interference predictor}
We propose an interference predictor that integrates heavy-tailed statistics, along with functional and covariance modeling, within a Bayesian framework, providing robustness to sporadic interference through \gls{SPTPR}. Specifically, the dynamic and measurement model is approximated as follows:
\begin{subequations}\label{eq:dssm}
\begin{align}
    &\boldsymbol{\iota}_n(t) \approx  \mathcal{STP}^{F}_{\mu}( {\boldsymbol{\iota}}_n(t-1)) + \hat{\boldsymbol{\beta}}(t) \label{eq:dssma}, \\
    &\mathbf{y}_{n}(t) \approx \mathcal{STP}^{G}_{\mu}({\boldsymbol{\iota}}_n(t)) + \hat{\mathbf{u}}(t) \label{eq:dssmb},
\end{align}
\end{subequations}
where, noise characterized by $ \hat{\boldsymbol{\beta}}(t) \sim \mathcal{TP}(0,\mathcal{STP}^{F}_{\Sigma}({\boldsymbol{\iota}}_n(t-1)),\nu^{F})$ and 
\mbox{$ \hat{\mathbf{u}}(t) \sim  \mathcal{TP}(0,\mathcal{STP}^{G}_{\Sigma}( {\boldsymbol{\iota}}_n(t)),\nu^{G})$}. The ideal outcome is for  $\mathcal{STP}^{F}_{\mu}$ and $\mathcal{STP}^{G}_{\mu}$ to approach to $F(\cdot)$ and $G(\cdot)$ and for $\mathcal{STP}^{F}_{\Sigma}$ and $\mathcal{STP}^{G}_{\Sigma}$ is approach to $\mathbf{C}_F$ and $\mathbf{C}_G$. For brevity, we define \(\mathcal{STP}^{F}_{\mu} = [\mathcal{STP}^{f_{1}}_{\mu}, \dotsc, \mathcal{STP}^{f_{M}}_{\mu}] \in \mathbb{R}^{1 \times M}\) and the diagonally loaded matrix \(\hat{\mathbf{C}}_F =\text{diag}[\mathcal{STP}^{f_{1}}_{\Sigma},\dotsc, \mathcal{STP}^{f_{M}}_{\Sigma}] \in \mathbb{R}^{M \times M}\), with analogous definitions applied to the measurement model. This covariance modeling allows flexibility to incorporate intra-\gls{SN} interference by updating the off-diagonal elements of the matrix. However, in this work, we assume minimal or negligible intra-\gls{SN} interference, owing to the use of an orthogonal scheduler as discussed in \cref{sec:systemmodel}. Furthermore, unlike conventional architectures that include a dequantizer to recover \gls{eSNR} values from \gls{CQI} reports, our proposed technique omits it, as the \gls{SPTPR} directly infers interference from \gls{CQI} based on~\cref{eq:dssmb}. Given the heavy-tailed and nonlinear nature of both \cref{eq:dssma} and \cref{eq:dssmb}, we employ the \gls{MUKF}, which uses the modified unscented transform to accurately propagate mean and covariance through the nonlinear dynamics, rather than relying on linear approximations as in \cite{pramesh_kalmam}. To efficiently handle heavy-tailed Student-t estimates, we adopt the sigma-point based on~\cite{ukf_huang}, which preserves key distributional characteristics for sigma-points and weights as:
\begin{equation*}
\resizebox{\linewidth}{!}{$
\left\{
\begin{aligned}
&\mathbf{x}_{m}^{i} = \mu_{m},~ 
\omega_{m}^{i,\mu} = \frac{\kappa}{s+\kappa},~  
 \omega_{{m}}^{i,\Sigma} = \frac{\kappa}{s+\kappa} + \left(1 - \alpha^2 + \beta\right),~  i = 0 \\
&\mathbf{x}_{m}^{i} = \mu_{m} + \sqrt{\frac{\nu(s+\kappa)}{\nu-2}}\sqrt{\boldsymbol{\Sigma}}  \mathbf{e}_i,~ \omega_{m}^{i,\mu} = \omega_{{m}}^{i,\Sigma} = \frac{0.5}{s+\kappa}
~  i = 1, \cdots, s \\
&\mathbf{x}_{m}^{i,\Sigma} = \mu_{m} - \sqrt{\frac{\nu(s+\kappa)}{\nu-2}}\sqrt{\boldsymbol{\Sigma}}  \mathbf{e}_i,~\omega_{m}^{i,\mu} = \omega_{{m}}^{i,\Sigma} = \frac{0.5}{s+\kappa}
~ i = s+1, \cdots, 2s
\end{aligned}
\right.
$}
\end{equation*}
where $\kappa, \alpha ,\beta$ are the parameters of the \gls{UT}, $\sqrt{\boldsymbol{\Sigma}}$ is the square-root matrix of covariance matrix, and $\mathbf{e}_i$ denotes the $i$th column vector of a unit matrix for each \gls{SA} with $2s$ sigma points. The terms $\mu_{m} := \mathcal{STP}^{f_{m}}_{\mu_{m}}$ (dynamic model) and $\mathcal{STP}^{g_{m}}_{\mu_{m}}$ (measurement model) are defined accordingly. The sigma points $\mathbf{x}_m^i$ are stacked column-wise into $\mathbf{X}_n \in \mathbb{R}^{M \times (2s+1)}$, the mean weights $\omega_{m}^{i,\mu}$ into $\mathbf{W}_m \in \mathbb{R}^{M \times (2s+1)}$, and the covariance weights $\omega_{m}^{i,\Sigma}$ into $\mathbf{W}_c \in \mathbb{R}^{M \times (2s+1)}$. To ensure robust and accurate state estimation with a theoretically exact update despite the delay, the delayed \gls{CQI} must be time-aligned with the predicted interference, as summarized in \cref{algo:gpukf}. For computationally efficient delay compensation with total delay of $d$, we retrieve the buffered interference estimate at \(\hat{\boldsymbol{\iota}}_{n}(t-d-1)\)  and propagate it using the dynamic model to obtain \(\hat{\boldsymbol{\iota}}_{n}(t-d)\)~(lines 1–5). Following this, the delayed \gls{CQI} measurement \(\mathbf{y}_{n}(t)\), which reflects the interference at time \(t-d\), is used to correct the previous state estimate through a measurement update (lines 6–15), resulting in the updated state \(\hat{\boldsymbol{\iota}}_{n}(t-d)\). To compensate for the \gls{CQI} acquisition delay and predict the interference estimate at time \(t\), i.e. \(\hat{\boldsymbol{\iota}}_{n}(t)\), this updated state \(\hat{\boldsymbol{\iota}}_{n}(t-d)\) is further propagated using the dynamic model (lines 16–20). This framework flexibly aligns with the current time, as shown in \cref{fig:proposed_LA} (a), by storing only $d$ predicted interferences for exact updates based on delayed-\gls{CQI}.

\begin{algorithm}[!t]
\caption{SPTPR-MUKF($\hat{\boldsymbol{\iota}}_n(t-d-1), \boldsymbol{\Sigma}(t-d-1), \mathbf{y}_n(t)$)}
\label{algo:gpukf}
\begin{algorithmic}[1]
\Statex \hspace*{-1em}\textbf{MUKF prediction step:}
\State Generate $\bar{\mathbf{X}}_n(t\!-\!d\!-\!1)$ with $\hat{\boldsymbol{\iota}}_n(t\!-\!d\!-\!1)$
\State Calculate $\bar{\mathbf{X}}_n(t\!-\!d), \hat{\mathbf{C}}_{F}(t\!-\!d)$ using (7a)
\State $\tilde{\boldsymbol{\iota}}_n(t\!-\!d) \leftarrow \sum_{j=0}^{2s} \mathbf{W}_m[:,j] \odot \bar{\mathbf{X}}_n[:,j](t\!-\!d)$
\State $\hat{\boldsymbol{\Sigma}}(t\!-\!d) \leftarrow \sum_{j=0}^{2s} \mathbf{W}_c[:,j] \odot 
\left( \bar{\mathbf{X}}_n[:,j](t\!-\!d) - \tilde{\boldsymbol{\iota}}_n(t\!-\!d) \right)$
\Statex \hspace*{2em}$
\left( \bar{\mathbf{X}}_n[:,j](t\!-\!d) - \tilde{\boldsymbol{\iota}}_n(t\!-\!d) \right)^\top+ \hat{\mathbf{C}}_{F}(t\!-\!d)$
\Statex\hspace*{-1.25em}\textbf{MUKF update step:}
\State Generate $\hat{\mathbf{X}}_n(t\!-\!d)$ centered at $\tilde{\boldsymbol{\iota}}_n(t\!-\!d)$
\State Calculate $\hat{\boldsymbol{\Gamma}}_n(t\!-\!d), \hat{\mathbf{C}}_{G}(t\!-\!d)$ using (7b)
\State $\hat{\mathbf{y}}_n(t\!-\!d) \leftarrow \sum_{j=0}^{2s} \mathbf{W}_m[:,j] \odot \hat{\boldsymbol{\Gamma}}_n[:,j](t\!-\!d)$
\State $\mathbf{S}(t\!-\!d) \leftarrow \sum_{j=0}^{2s} \mathbf{W}_c[:,j] \odot 
\left( \hat{\boldsymbol{\Gamma}}_n[:,j](t\!-\!d) - \hat{\mathbf{y}}_n(t\!-\!d) \right)$
\Statex \hspace*{1em}$\left( \hat{\boldsymbol{\Gamma}}_n[:,j](t\!-\!d) - \hat{\mathbf{y}}_n(t\!-\!d) \right)^\top + \hat{\mathbf{C}}_{G}(t\!-\!d)$
\State $\boldsymbol{\Sigma}_{\iota y}(t\!-\!d) \leftarrow \sum_{j=0}^{2s} \mathbf{W}_c[:,j] \odot $
\Statex $\left( \hat{\mathbf{X}}_n[:,j](t\!-\!d) - \hat{\boldsymbol{\iota}}_n(t\!-\!d) \right)\left( \hat{\boldsymbol{\Gamma}}_n[:,j](t\!-\!d) - \hat{\mathbf{y}}_n(t\!-\!d) \right)^\top$
\State $\mathbf{K}(t\!-\!d) \leftarrow \boldsymbol{\Sigma}_{\iota y}(t\!-\!d) \mathbf{S}(t\!-\!d)^{-1}$
\State $\hat{\boldsymbol{\iota}}_n(t\!-\!d) \leftarrow \tilde{\boldsymbol{\iota}}_n(t\!-\!d) + \mathbf{K}(t\!-\!d)\left(\mathbf{y}_n(t) - \hat{\mathbf{y}}_n(t\!-\!d)\right)$
\State $\boldsymbol{\Sigma}(t\!-\!d) \leftarrow \hat{\boldsymbol{\Sigma}}(t\!-\!d) - \mathbf{K}(t\!-\!d) \mathbf{S}(t\!-\!d) \mathbf{K}(t\!-\!d)^\top$
\Statex \hspace*{-1.4em}\textbf{Delay-compensation step:}
\For{$l = 0,\dotsc,d-1$} 
    \State Calculate $\hat{\mathbf{X}}_n(t\!-\!d\!+\!l\!+\!1)$ with $\hat{\boldsymbol{\iota}}_n(t\!-\!d+l)$  using (7a)
    \State $\hat{\boldsymbol{\iota}}_n(t\!-\!d\!+\!l\!+\!1) \leftarrow \sum_{j=0}^{2s} \mathbf{W}_m[:,j] \odot \hat{\mathbf{X}}_n[:,j](t\!-\!d\!+\!l\!+\!1)$
    \State Store $\hat{\boldsymbol{\iota}}_n(t\!-\!d\!+\!l\!+\!1)$, $\boldsymbol{\Sigma}(t\!-\!d\!+\!l\!+\!1)$ in buffer
\EndFor
\State \textbf{return} $\hat{\boldsymbol{\iota}}_n(t)$, $\boldsymbol{\Sigma}(t)$
\end{algorithmic}
\end{algorithm}

Overall, our proposed predictive \gls{LA}, which compensates for delay, is illustrated in Fig.~\ref{fig:proposed_LA}, with the \gls{IPV} prediction module highlighted in orange.  By predicting interference directly from \gls{CQI}, this approach removes the need for a dequantizer and avoids \gls{HARQ}-based adjustments, which are impractical under 0.1–1 ms latency. Moreover, this method enables proactive SINR adaptation for efficient \gls{LA} for densely deployed interference-prone \glspl{SN}.

\section{Numerical Results}
\label{sec:simulations}
\subsection{Baselines and Performance Metrics}
We evaluate the proposed \gls{SPTPR}-\gls{MUKF} against baseline predictors as follows:
\begin{enumerate}[label=\arabic*)]
    \item \textit{Genie \gls{LA}:} This baseline performs \gls{MCS} selection based on the target \gls{BLER} $\Bar{\varepsilon}$ under the assumption that a \gls{SA}'s \gls{SINR} is perfectly available without delay at the \gls{SNC}. Note that Genie \gls{LA} is impractical and is considered here as a sort of upper bound for \gls{MCS} selection~\cite{pramesh_kalmam}.
    \item \textit{\Gls{LSTM} Predictor in \gls{SA}:} In this supervised learning-based baseline, a pre-trained \gls{LSTM} model predicts \glspl{IPV} at each $t$ based on a \glspl{IPV} of predefined window as its input for $M$ \glspl{SA} \cite{gautam2023co}.
    \item Conventional \textit{\Gls{MA} Predictor} \cite{pocovi2017mac}.
\end{enumerate}

\refstepcounter{figure}
\begin{figure*}[t]
\centering
\captionsetup[subfloat]{labelformat=empty}
\subfloat[Fig. 2a: Achieved BLER for single \gls{SA}.]{\label{fig:Fig2a} \includegraphics[width=0.28\linewidth]{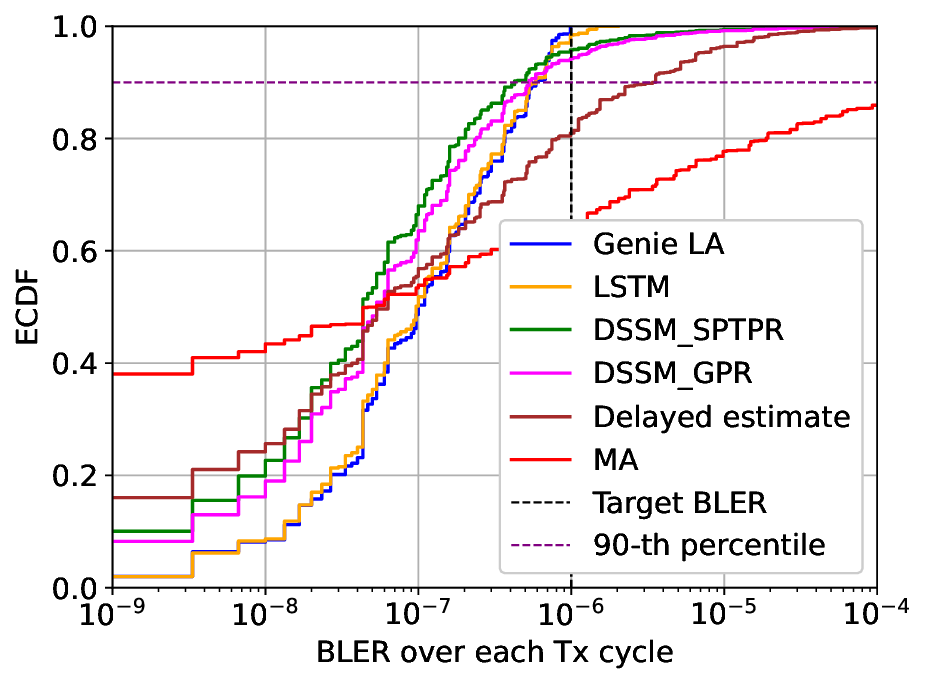}}%
\subfloat[Fig. 2b: Scaling number of interferers.]{\label{fig:Fig2c} \includegraphics[width=0.28\linewidth]{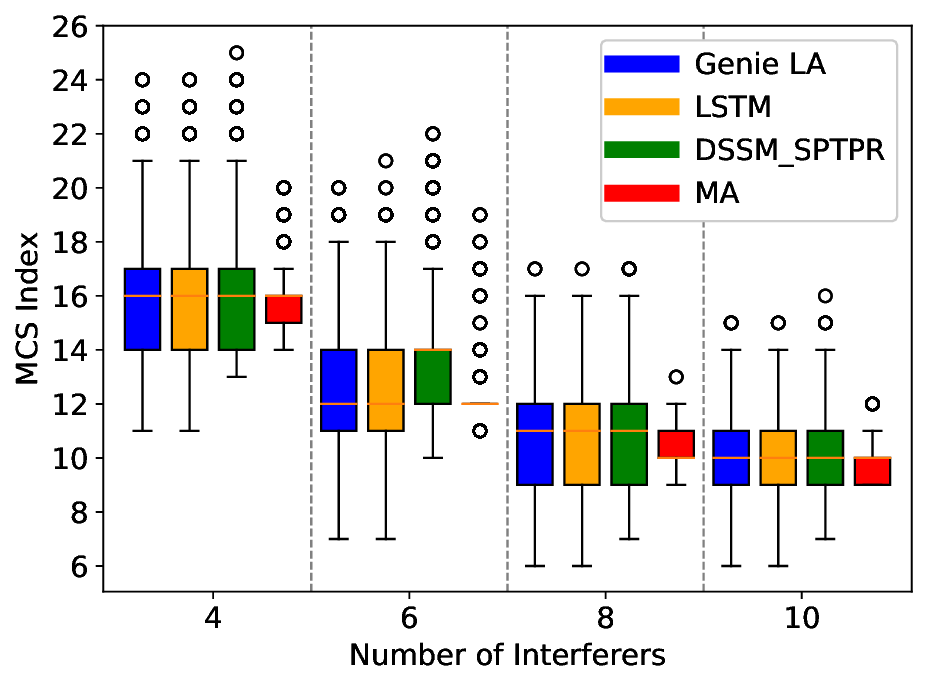}}
\subfloat[Fig. 2c: Achieved BLER for multi-SAs.]{\label{fig:Fig2b} \includegraphics[width=0.28\linewidth]{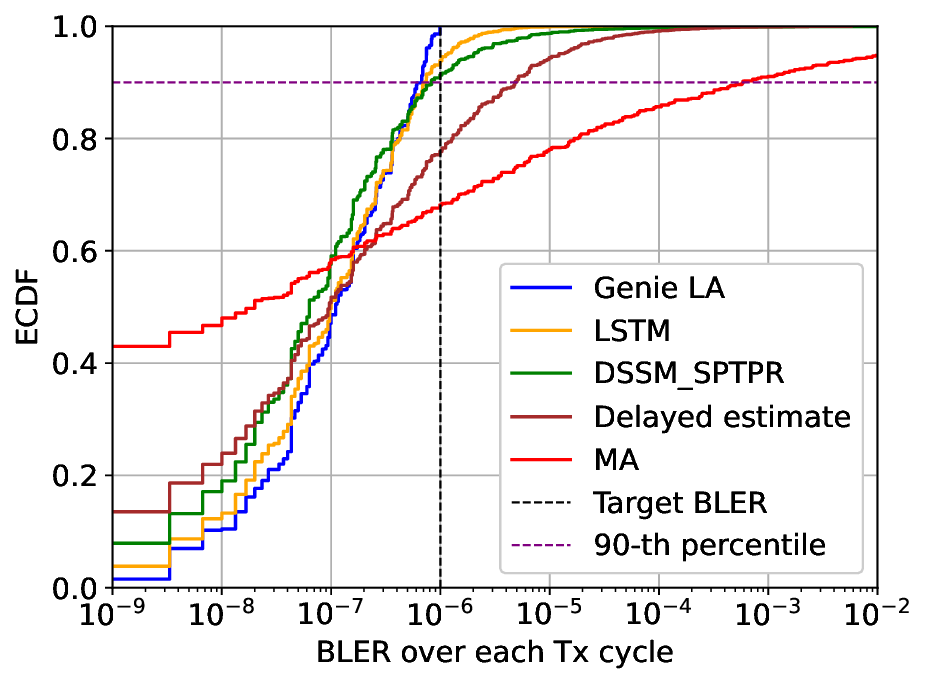}}%
\vspace{-0.2in}
\end{figure*}

The performance is evaluated using the following metrics:
\subsubsection{Instantaneous \gls{SE}}
It captures the stochastic nature of transmission success in a \gls{SN}, and is defined as:   \begin{equation}
    {\text{SE}}_{\text{inst}}(t) = B(1 - \varepsilon(t)) \cdot R,
\end{equation}
where \(B(t)\) is a Bernoulli random variable with success probability \(1 - \varepsilon(t)\), and \(\varepsilon(t)\) denotes the BLER. With probability \(1 - \varepsilon(t)\), the system achieves spectral efficiency \(R\); otherwise, it achieves zero.  
\subsubsection{Target vs achieved \gls{BLER}}
It quantifies the gap between the target \gls{BLER} and the achieved \gls{BLER}. \gls{MCS} is selected using the target \gls{BLER} and predicted \gls{SINR}, while the achieved \gls{BLER} is calculated using the selected \gls{MCS} for true \gls{SINR}. 

\subsection{Simulation Setup and Results}
We evaluate the proposed approach for \gls{LA} improvements through numerical simulations and compare it against baseline methods. We consider \(N = 20\) \glspl{SN} deployed in a \(25 \times 25~\text{m}^2\) factory hall, with each \gls{SN} with either \(M = 1\) for the single \gls{SA} case or \(M = 4\) \glspl{SA} for the multi-\gls{SA} case. Each \gls{SN} operates at 6 GHz across 4 available sub-bands. Interference and channel conditions are modeled according to the 3GPP InF-DL specification~\cite{3gpp2020study} as detailed in \cref{sec:systemmodel}. The \glspl{SN} move at 2 m/s in random directions while avoiding collisions, resulting in a Doppler frequency of 80 Hz. We assume a \gls{SN}  transmits 160-bit packets at 0 dBm power, with a 1 ms \gls{Tx} cycle and an inherent protocol delay of $d = 2$ ms. We calculate wideband \gls{CQI} with \(|\zeta| = 29\) \gls{MCS} levels based on~\cite{3gpp2018physical}. Fig. 2a shows that relying on delayed, dequantized \gls{SINR} estimates with a \(d = 2\) ms lag performs better than \gls{MA} but leads to suboptimal \gls{MCS} selection and degraded \gls{BLER}, underscoring the need for protocol-delay compensation. 
In contrast, the supervised \gls{LSTM} baseline, trained for 300 epochs on sliding windows of 30 \gls{IPV}, performs rolling one-step-ahead predictions to compensate for the \gls{CQI} delay at \gls{SA}. Its gated temporal modeling enables \gls{BLER} performance close to that of the Genie \gls{LA}; it critically relies on perfect noise-free \gls{IPV} known at \gls{SN} . In contrast, by utilizing outdated \gls{CQI} within our proposed framework, the \gls{GPR} with the \gls{UKF} model achieves near-optimal performance without requiring additional signaling overhead. Both dynamic and measurement models are trained over 300 epochs using a squared exponential kernel $k(\cdot)$~\cite{rasmussen2010gaussian} with learning rates of 0.04 and 0.01. For the \gls{SPTPR}, \(L_E = 300\) inducing points are empirically selected from \(L = 1000\) training samples, enabling a computationally scalable approach to inference. The predicted interference is refined recursively with observed \gls{CQI} utilizing \gls{MUKF}, which is configured with \(\kappa = 0.96\) and degrees of freedom \(\nu = 3\), selected heuristically. The proposed \gls{SPTPR}-\gls{MUKF}, which leverages a heavy-tailed prior, delivers near-optimal performance, achieving the 90th percentile \gls{BLER} target, while being robust to outliers and reducing computational cost by over 10$\times$. Fig. 2b illustrates its scalability under increasing interference for single \gls{SA}. As expected, the distribution of \gls{MCS} orders shifts lower, reducing spectral efficiency. Meanwhile, the distribution of \gls{MCS} based on our proposed technique remains close to that of the Genie LA. Moreover, Fig. 2c illustrates results for multi-\gls{SA} scenarios, where the proposed method achieves the 90th percentile target BLER as well. 
Table \ref{tab:se_performance} then shows that \gls{SPTPR}-\gls{MUKF} achieves competitive spectral efficiency across all \glspl{SA}. Table \ref{tab:delay_impact} demonstrates that our proposed framework, \gls{SPTPR}-\gls{MUKF}, effectively mitigates the impact of protocol delay by substantially improving performance and achieving the 90th percentile BLER target of $10^{-6}$, even under a delay of 10 ms in the presence of dynamic interference. It achieves performance comparable to LSTM utilizing available channel state information, i.e., CQI only, without requiring additional signaling overhead for transmission of interference estimates on every \gls{Tx} cycle at the \gls{SA}.

\begin{table}[ht]
\centering
\caption{Average $\text{SE}_{\text{inst}}$ with $d = 2$ ms for multi-\glspl{SA}.}
\label{tab:se_performance}
\begin{tabular}{|l|l|l|l|l|}
\hline
\textbf{Techniques} & $m_1$ & $m_2$ & $m_3$ & $m_4$ \\ \hline
Genie LA             & 2.69  & 2.53  & 2.39  & 1.93  \\ \hline
LSTM                & 2.68  & 2.53  & 2.38  & 1.92  \\ \hline
DSSM-SPTPR          & 2.66  & 2.50  & 2.37  & 1.95  \\ \hline
\end{tabular}
\vspace{-0.15in}
\end{table}

\begin{table}[ht]
\centering
\caption{Achieved 90th-percentile BLER at various protocol delays, with a target BLER of $10^{-6}$.}
\label{tab:delay_impact}
\resizebox{0.5\textwidth}{!}{%
\begin{tabular}{|l|l|l|l|l|}
\hline
\textbf{Techniques}  & 2 ms & 4 ms & 8 ms & 10 ms \\ \hline
Delayed estimate     & \(1.54 \times 10^{-6}\) & \(5.19 \times 10^{-6}\) & \(38.8 \times 10^{-6}\) & \(130.6 \times 10^{-6}\) \\ \hline
LSTM                 & \(0.530 \times 10^{-6}\) & \(0.530 \times 10^{-6}\) & \(0.552 \times 10^{-6}\) & \(0.605 \times 10^{-6}\) \\ \hline
DSSM-SPTPR           & \(0.260 \times 10^{-6}\) & \(0.514 \times 10^{-6}\) & \(0.828 \times 10^{-6}\) & \(0.988 \times 10^{-6}\) \\ \hline
\end{tabular}%
}
\vspace{-0.15in}
\end{table}



\section{Conclusions}

We proposed SPTPR-MUKF, a semi-supervised interference prediction method for link adaptation in industrial subnetworks, leveraging CQI to infer heavy-tailed interference via a sparse Student-t process within a \gls{vDSSM}. The approach achieves near-optimal performance with over 10× lower complexity, while remaining robust to protocol delays, outliers, and scalable under dense deployments. Future work will explore real-time deployment, incorporating random CQI acquisition delays and generalization across diverse deployment scenarios.


\begin{thebibliography}{10}
\providecommand{\url}[1]{#1}
\csname url@samestyle\endcsname
\providecommand{\newblock}{\relax}
\providecommand{\bibinfo}[2]{#2}
\providecommand{\BIBentrySTDinterwordspacing}{\spaceskip=0pt\relax}
\providecommand{\BIBentryALTinterwordstretchfactor}{4}
\providecommand{\BIBentryALTinterwordspacing}{\spaceskip=\fontdimen2\font plus
\BIBentryALTinterwordstretchfactor\fontdimen3\font minus \fontdimen4\font\relax}
\providecommand{\BIBforeignlanguage}[2]{{%
\expandafter\ifx\csname l@#1\endcsname\relax
\typeout{** WARNING: IEEEtran.bst: No hyphenation pattern has been}%
\typeout{** loaded for the language `#1'. Using the pattern for}%
\typeout{** the default language instead.}%
\else
\language=\csname l@#1\endcsname
\fi
#2}}
\providecommand{\BIBdecl}{\relax}
\BIBdecl

\bibitem{brighente2020interference}
A.~Brighente, J.~Mohammadi \emph{et~al.}, ``{Interference Prediction for Low-Complexity Link Adaptation in Beyond 5G Ultra-Reliable Low-Latency Communications},'' \emph{IEEE Transactions on Wireless Communications}, vol.~21, no.~10, pp. 8403--8415, 2022.

\bibitem{berardinelli2021extreme}
G.~Berardinelli, P.~Baracca \emph{et~al.}, ``{Extreme Communication in 6G: Vision and Challenges for ‘in-X’ Subnetworks},'' \emph{IEEE Open Journal of the Communications Society}, vol.~2, pp. 2516--2535, 2021.

\bibitem{10024770}
X.~Ye, Y.~Yu, and L.~Fu, ``{Deep Reinforcement Learning Based Link Adaptation Technique for LTE/NR Systems},'' \emph{IEEE Transactions on Vehicular Technology}, vol.~72, no.~6, pp. 7364--7379, 2023.

\bibitem{pramesh_kalmam}
P.~Gautam, R.~S. B.~A. G \emph{et~al.}, ``{Dynamic Interference Prediction for In-X 6G Sub-Networks},'' in \emph{2025 14th International ITG Conference on Systems, Communications and Coding (SCC)}, 2025, pp. 1--6.

\bibitem{pocovi2017mac}
G.~Pocovi, B.~Soret \emph{et~al.}, ``{MAC layer enhancements for ultra-reliable low-latency communications in cellular networks},'' in \emph{2017 IEEE International Conference on Communications Workshops (ICC Workshops)}, 2017, pp. 1005--1010.

\bibitem{wei2024joint}
L.~Wei, Z.~Liu \emph{et~al.}, ``{Joint Model and Data-Driven Two-Stage Uplink Interference Prediction in URLLC Scenarios},'' in \emph{2024 IEEE Wireless Communications and Networking Conference (WCNC)}, 2024, pp. 1--6.

\bibitem{xu2013improving}
X.~Xu, M.~Ni \emph{et~al.}, ``{Improving QoS by predictive channel quality feedback for LTE},'' in \emph{21st International Conference on Software, Telecommunications and Computer Networks}, 2013, pp. 1--5.

\bibitem{clavier2020experimental}
L.~Clavier, T.~Pedersen, I.~Larrad, M.~Lauridsen, and M.~Egan, ``{Experimental evidence for heavy tailed interference in the IoT},'' \emph{IEEE Communications Letters}, vol.~25, no.~3, pp. 692--695, 2020.

\bibitem{3gpp2018physical}
{3GPP}, ``{Physical layer procedures for data (Release 16)},'' 3rd Generation Partnership Project, Tech. Rep. 38.214 v16.2.0, 2020.

\bibitem{gautam2023co}
P.~Gautam, M.~Vakilifard \emph{et~al.}, ``{Cooperative Interference Estimation Using LSTM-Based Federated Learning for In-X Subnetworks},'' in \emph{GLOBECOM 2023 - 2023 IEEE Global Communications Conference}, 2023, pp. 1338--1344.

\bibitem{rasmussen2010gaussian}
C.~K. Williams and C.~E. Rasmussen, \emph{{Gaussian processes for machine learning}}.\hskip 1em plus 0.5em minus 0.4em\relax MIT press Cambridge, MA, 2006, vol.~2, no.~3.

\bibitem{shah2014student}
A.~Shah, A.~Wilson, and Z.~Ghahramani, ``{Student-t processes as alternatives to Gaussian processes},'' in \emph{Artificial intelligence and statistics}.\hskip 1em plus 0.5em minus 0.4em\relax PMLR, 2014, pp. 877--885.

\bibitem{xu2024sparse}
J.~Xu and D.~Zeng, ``{Sparse Variational Student-t Processes},'' in \emph{Proceedings of the AAAI Conference on Artificial Intelligence}, vol.~38, no.~14, 2024, pp. 16\,156--16\,163.

\bibitem{ukf_huang}
Y.~Huang, Y.~Zhang \emph{et~al.}, ``Robust student’s t based nonlinear filter and smoother,'' \emph{IEEE Transactions on Aerospace and Electronic Systems}, vol.~52, no.~5, pp. 2586--2596, 2016.

\bibitem{3gpp2020study}
3GPP, ``{5G: Study on channel model for frequencies from 0.5 to 100 GHz (Release 16)},'' 3rd Generation Partnership Project, Tech. Rep. 38.901 v16.1.0, 2020.

\end{thebibliography}
\end{document}